\documentclass[usenatbib,letterpaper]{mnras}
\usepackage{amssymb}
\usepackage{color}
\usepackage{url}
\usepackage{graphicx}
\usepackage{times}
\usepackage[T1]{fontenc}
\usepackage{aecompl}
\usepackage{amssymb}
\usepackage{caption}

\defcitealias{Krolik:2015lr}{KH15}

\title[Apsidal precession, disc breaking and viscosity]{Apsidal precession, disc breaking and viscosity in warped discs}
\author[R. Nealon et al.]{Rebecca Nealon$^{1}$\thanks{rebecca.nealon@monash.edu}, Chris Nixon$^{2,3}$\thanks{Einstein Fellow}, Daniel J. Price$^{1}$ \& Andrew King$^{3}$\\
$^1$Monash Centre for Astrophysics (MoCA) and School of Physics and Astronomy, Monash University, Clayton VIC 3800, Australia\\
$^2$JILA, University of Colorado \& NIST, Boulder CO 80309-0440, USA\\
$^3$Department of Physics \& Astronomy, University of Leicester, Leicester LE1 7RH UK\\  
}
\pagerange{\pageref{firstpage}--\pageref{lastpage}} \pubyear{2015}

\begin{document}
\label{firstpage}
\maketitle

\begin{abstract}
We demonstrate the importance of general relativistic apsidal precession in warped black hole accretion discs by comparing three--dimensional smoothed particle hydrodynamic simulations in which this effect is first neglected, and then included. 
If apsidal precession is neglected, we confirm the results of an earlier magnetohydrodynamic simulation which made this assumption, showing that at least in this case the $\alpha$ viscosity model produces very similar results to those of simulations where angular momentum transport is due to the magnetorotational instability. Including apsidal precession significantly changes the predicted disc evolution. For moderately inclined discs thick enough that tilt is transported by bending waves, we find a disc tilt which is nonzero at the inner disc edge and oscillates with radius, consistent with published analytic results. For larger inclinations we find disc breaking. 
\end{abstract}


\section{Introduction}
\citet{bardeen_petterson_1975} first studied accretion discs misaligned with the spin of a central black hole. The Lense-Thirring effect causes the disc to warp, so that the inclination between the local disc plane and the black hole spin varies as a function of radius. \citet{pap_pringle_1983} identified two regimes that govern the propagation of warps, determined by comparing the dimensionless viscosity parameter $\alpha$ \citep{shakura_sunyaev} to the aspect ratio $H/R$, where $H$ is the disc scale height and $R$ the radius. When $\alpha \gtrsim H/R$, warp evolution is described by a diffusion equation. In constrast when $\alpha \lesssim H/R$, warps propagate as bending waves that travel at half the sound speed \citep{Papaloizou:1995pn}. \citet{Papaloizou:1995lr} showed that the disc can warp significantly when the sound crossing time is longer than the induced precession time. See \cite{Nixon:2015fk} for a recent review of warped disc dynamics.

In the bending wave regime, \citet{i_and_i_1997} and \citet*{lubow2002evolution} showed that it is possible for a steady--state accretion disc to be misaligned at the inner edge, with its plane oscillating with radius close to the black hole.  \citet{nelson_pap_2000} were the first to investigate this behaviour using numerical simulations, and recently non--zero inner edge tilts \citep[e.g. $10^{\circ}$,][]{Zhuravlev:2014fk} and oscillations of the tilt with radius \citep*[e.g. $3^{\circ}$ and $15^{\circ}$,][]{Nealon:2015fk} have been confirmed. These results suggest that the oscillatory tilt behaviour is only possible when general relativistic (GR) effects near the black hole -- specifically the apsidal (Einstein) precession -- are modelled accurately. In contrast, \citet[][hereafter KH15]{Krolik:2015lr} state that the effect of the apsidal precession on the disc is `at most a minor physical influence' and so do not include it in their simulation of an accretion disc in the bending wave regime.  Their simulations show alignment at the inner edge, and no oscillations, contrary to these earlier results. This suggests that neglecting apsidal precession may produce significantly different disc behaviour near the black hole. Here we try to clarify this and other questions.

When the internal torques in the disc (from pressure and viscosity) are too weak to communicate an external forced precession (e.g. Lense--Thirring precession from a spinning black hole) the disc can break or tear into discrete planes. This is predicted by analytic calculations of \citet{Ogilvie:1999lr} and was found numerically by e.g. \citet{nixon_king_2011} and \citet{nixon_et_al_2012}. Recent smoothed particle hydrodynamic (SPH) simulations of warped accretion discs have explored this behaviour  (e.g. \citealt{nixon_et_al_2012b}; \citealt*{nixon_2013}; \citealt{Nealon:2015fk,Dogan:2015rt,Aly:2015tf}). Similar evolution has also been seen in a grid based code \citep{Fragner:2010rt}. In addition to this, features observed in the accretion disc around HD142527 appear to show a direct example of disc breaking \citep{Casassus:2015yu}. 

No MHD simulation has yet found disc breaking \citep[e.g.][]{Fragile:2007uq,Fragile:2008qy,Sorathia:2013fk,Morales-Teixeira:2014lr,Krolik:2015lr}. This is expected, because the analytic criteria (see below) predict that breaking occurs in thin discs with significant misalignments. MHD simulations are currently restricted to thick discs (the thinnest has $H/R \sim 0.08$, \citealt{Morales-Teixeira:2014lr}) with small misalignments ($\lesssim 20^{\circ}$).
\citetalias{Krolik:2015lr} simulated a disc which they argued should break according to the diffusive--regime criterion of \cite{nixon_et_al_2012b}. 
%
%
 In fact the disc \citetalias{Krolik:2015lr} simulate is in the wavelike regime, and when using the breaking criterion for this case \citep{Nealon:2015fk}, \citetalias{Krolik:2015lr} indeed also concluded it should not break.

Here we report a simulation with the same initial disc setup as \citetalias{Krolik:2015lr}. This is a purely hydrodynamic calculation with an $\alpha$ viscosity. We also find it does not break -- in agreement with the wavelike criterion \citep{Nealon:2015fk}. Our results are close to those of \citetalias{Krolik:2015lr}, suggesting that the use of the full MHD equations does not strongly affect the results in this case.

Grid-based simulations have also attempted to quantify the isotropy of $\alpha$ by measuring the local instantaneous stress as a function of position in simulations of warped accretion discs \citep{Sorathia:2013fk,Morales-Teixeira:2014lr}. These measurements show small scale ($\ll H$) variations in the stress, prompting  the authors to suggest that the $\alpha$ model may not give valid results. However, as discussed in \citet{Nixon:2015lr}, $\alpha$ is an averaged quantity, so it is not clear how these local measurements can be used to draw conclusions about the $\alpha$ model. We examine this by comparing our simulation with a fixed $\alpha$ to the MHD simulation of \citetalias{Krolik:2015lr}.

We check the importance of apsidal precession by performing 3D SPH simulations. We start with a disc that matches the simulation conducted by \citetalias{Krolik:2015lr}, and, as there, apsidal precession is neglected. In a second simulation of this same disc we then include the effects of apsidal precession and compare the disc evolution in these two cases. We perform a third simulation, this time with a higher disc inclination, and find that the disc breaks.

\begin{figure}
\includegraphics [width=0.9\columnwidth] {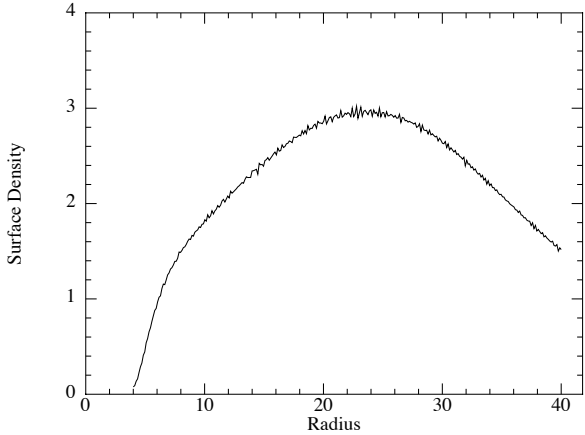}
\caption{Surface density profile of the initial condition used for all simulations, chosen to match \citetalias{Krolik:2015lr}.}
\label{fig:relaxed_sigma}
\end{figure}

\section{Simulations}
\label{section:numerical_setup}
We use 3D simulations to investigate the effect of apsidal precession on the evolution of accretion discs. We use the SPH code \textsc{Phantom} \citep{lodato_2010,price_2010,Price:2012fj}. This code has been used in numerous previous investigations of warped accretion discs in the diffusive regime $\alpha \lesssim H/R$
\citep[e.g.][]{lodato_2010,nixon_2012,nixon_et_al_2012b,nixon_et_al_2012,nixon_2013,Dogan:2015rt}. Numerical tests also show that it can accurately capture bending waves \citep{facchini_2013,Nealon:2015fk}.

\subsection{Disc parameters}
We match our initial disc setup to the recent simulation by \citetalias{Krolik:2015lr} for the purpose of comparison. We thus adopt an isothermal disc with sound speed set to $c_s(R)=c_{s,in}R^{-q}$ and a surface density profile $\Sigma(R) = \Sigma_{in}R^{-p}$. We conducted simulations with $10^5$, $10^6$ and $10^7$ particles to check for convergence, and the results are presented with $10^7$ particles.

As in \citetalias{Krolik:2015lr}, the disc spans $R_{in} = 4$ to $R_{out}=40$, with aspect ratio  $H/R_{in}=0.06$ and central density $\Sigma_{in}=0.1504$ at the inner edge. To match the surface density profile after relaxation and the scale height, we set $p=-1$ and $q=0$. As we model the physical viscosity using the method described in \citet{lodato_2010}, this results in an $\alpha$ viscosity that varies slightly with $R$.

\citetalias{Krolik:2015lr} confirm their disc is in the bending wave regime by measuring both waves travelling at $0.5c_s$ and the viscosity parameter $\alpha$ directly from their simulation. We seek to enforce this in our simulations by setting $\alpha$ such that $\alpha\sim0.12$ at $R=5$ and $\alpha\sim0.049$ at $R=10$, very similar to the values of $0.1$ and $0.05$ measured at the same locations in \citetalias{Krolik:2015lr}. With this description, $H/R \lesssim \alpha$ when $R\gtrsim 7$, suggesting that most of the disc will be in the bending wave regime.

The spin parameter $a$ is set from the requirement that the Lense-Thirring precession frequency is $1/15\times $ the orbital frequency at $R=10$ \citepalias{Krolik:2015lr}. As discussed in \citet{Sorathia:2013fk}, this constraint is made for numerical convenience as it can only be achieved with a non-physical value of $a\geq1.05$. In these simulations, we chose the spin to be the maximum physical value of $a=1.0$ (although in reality the spin is likely to saturate at a value somewhat lower than this). We note that this (minor) discrepancy appears in our results in that the black hole torque appears to affect the disc more slowly than observed in \citetalias{Krolik:2015lr}.

\begin{figure*}
\includegraphics [width=0.9\columnwidth] {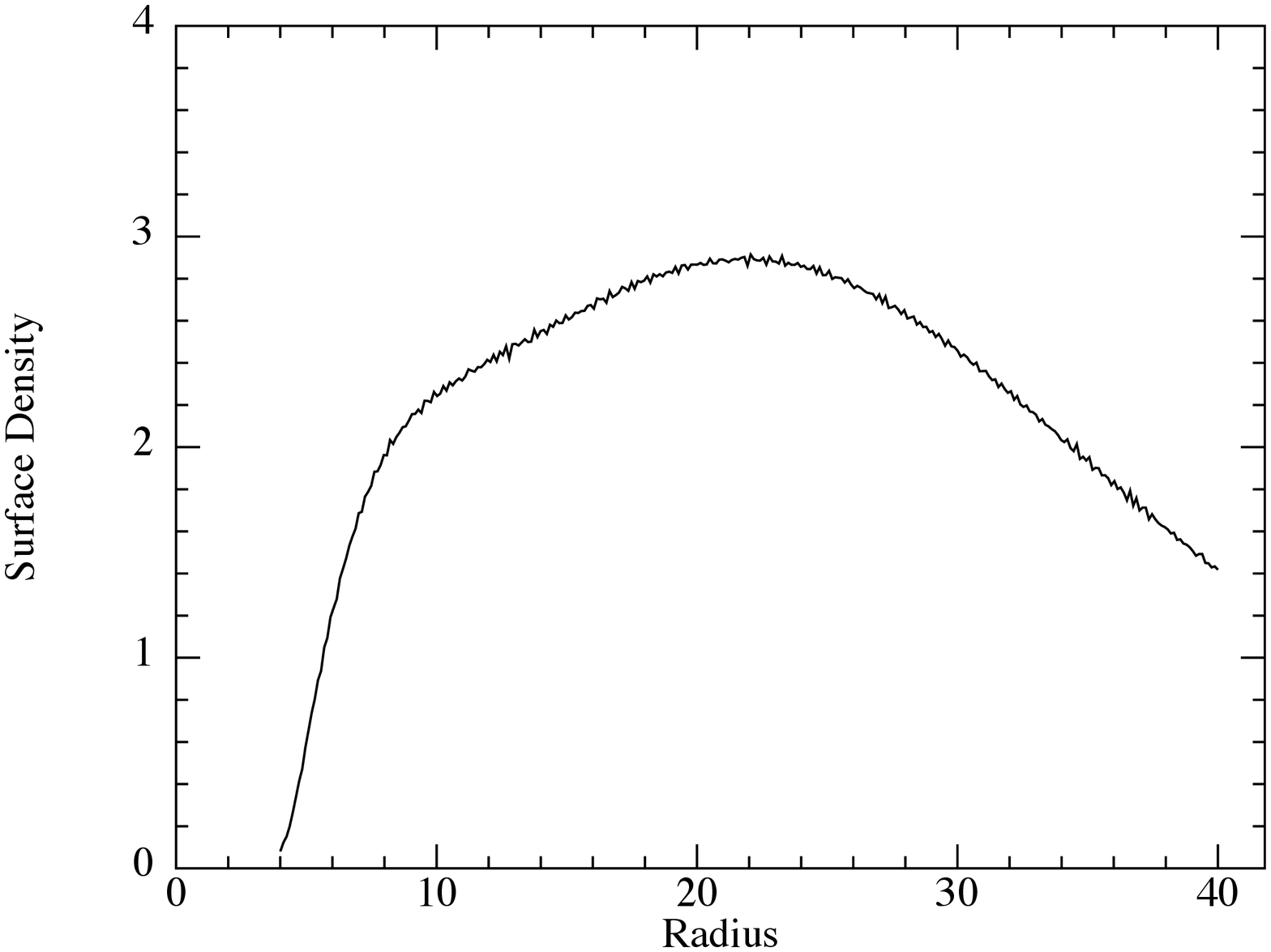} \hspace{1cm}
\includegraphics [width=0.9\columnwidth] {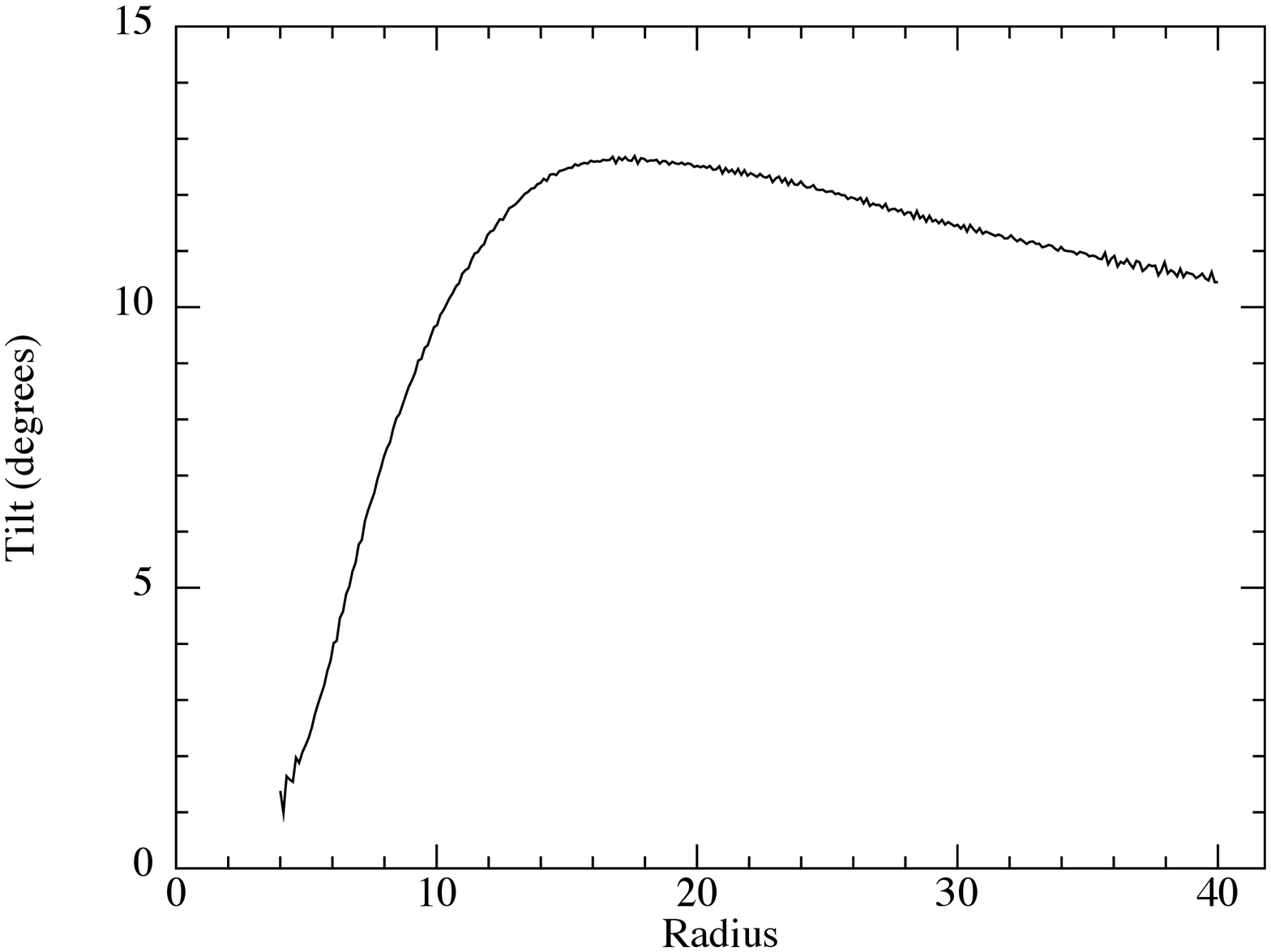}
\caption{Surface density (left) and tilt (right) after 25 orbits when apsidal precession is neglected.}
\label{fig:lt_only}
\end{figure*}

\begin{figure*}
\centering
\begin{minipage}{.47\textwidth}
  \centering
  \includegraphics[width=0.95\columnwidth]{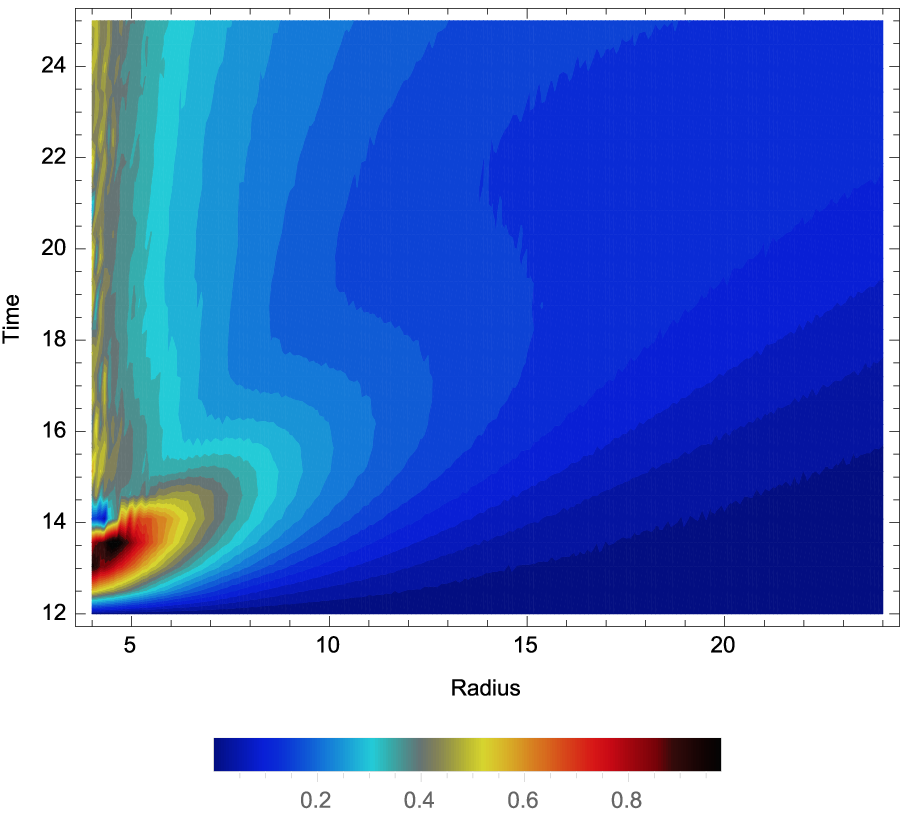}
  \captionof{figure}{Twist, measured in radians$/\pi$, as a function of time and radius ignoring apsidal precession. This figure, assuming the same precession physics, should be compared with Figure 2 of \citetalias{Krolik:2015lr}.}
  \label{fig:twist_lt}
\end{minipage}\hfill
\begin{minipage}{.47\textwidth}
  \centering
  \includegraphics[width=0.95\columnwidth]{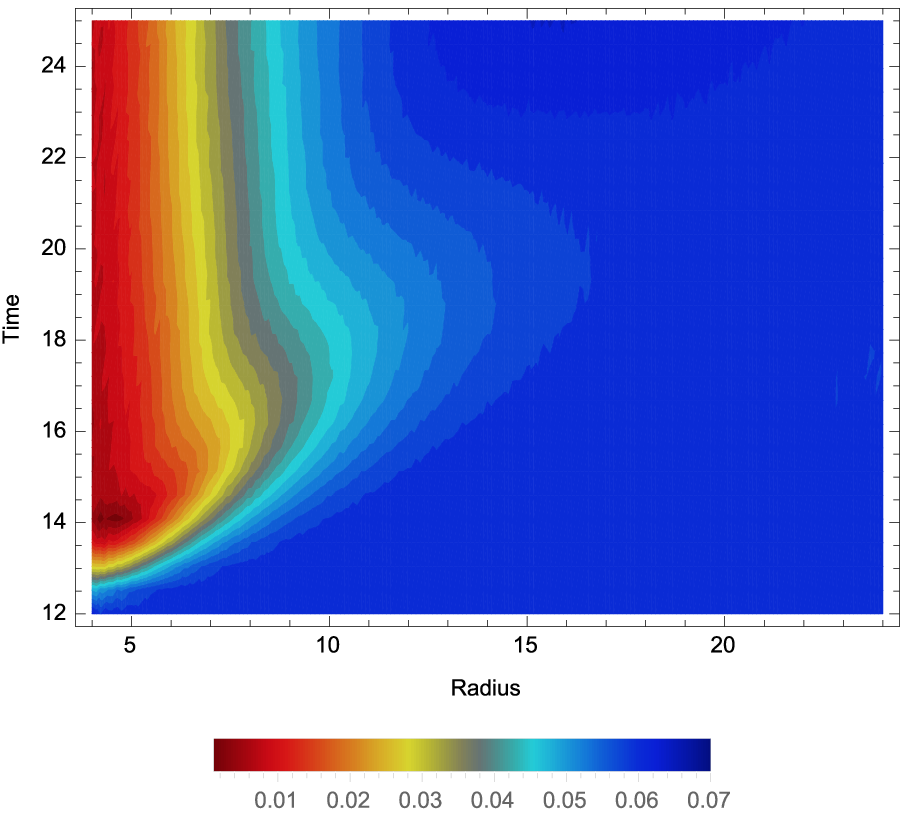}
  \captionof{figure}{Tilt, measured in radians$/\pi$, as a function of time and radius ignoring apsidal precession. This figure, assuming the same precession physics, should be compared with Figure 3 of \citetalias{Krolik:2015lr}. Note \citetalias{Krolik:2015lr} define the tilt to be the negative of our definition.}
  \label{fig:tilt_lt}
\end{minipage}
\end{figure*}

\subsection{Initial conditions}
\label{subsection:ic}
We arrange the particles in the disc using a Monte Carlo placement method, with the disc aligned to the black hole spin. The disc is then allowed to relax for $12.5$ orbits at $R=10$ in this plane \citepalias[12.4 orbits were conducted in][]{Krolik:2015lr}. Figure~\ref{fig:relaxed_sigma} shows the surface density profile at this time, showing that most of the mass is in the outer regions. Our disc shows the same features as \citetalias{Krolik:2015lr} (see their Figure 1) but has a slightly lower mass. We use this disc as the initial condition for each of our subsequent simulations, to allow direct comparison with \citetalias{Krolik:2015lr}. We produce the same surface density profile for the discs that include apsidal precession by repeating this process with the full effective GR potential.

The disc inclination $\beta(R)$  is the angle between the local angular momentum vector of the disc and the $z$ axis (defined by the black hole spin). For each simulation, the particles are rotated through a constant inclination angle $\beta$, so the disc is tilted but not warped. The black hole torque is then applied using either the Keplerian or effective GR potential as appropriate \citep[for details see][]{Nealon:2015fk}. We use units such that $G=M=1$ and $R_g=GM/c^2=1$ and define our timescale with orbits at $R=10$.

\section{Results}
\label{section:results}
We conduct two simulations with the only difference being the modelling of the precession frequencies \citep{Nealon:2015fk}. Both simulations implement Lense--Thirring precession, but the first simulation uses a Keplerian potential (as in \citetalias{Krolik:2015lr})
while the second simulation includes the effects of apsidal precession by using the effective GR potential. The discs are inclined at $\beta=12^{\circ}$ and run for an additional $12.5$ orbits from the initial condition described in Section~\ref{subsection:ic}, as described in \citetalias{Krolik:2015lr}.

\begin{figure*}
\includegraphics [width=0.9\columnwidth] {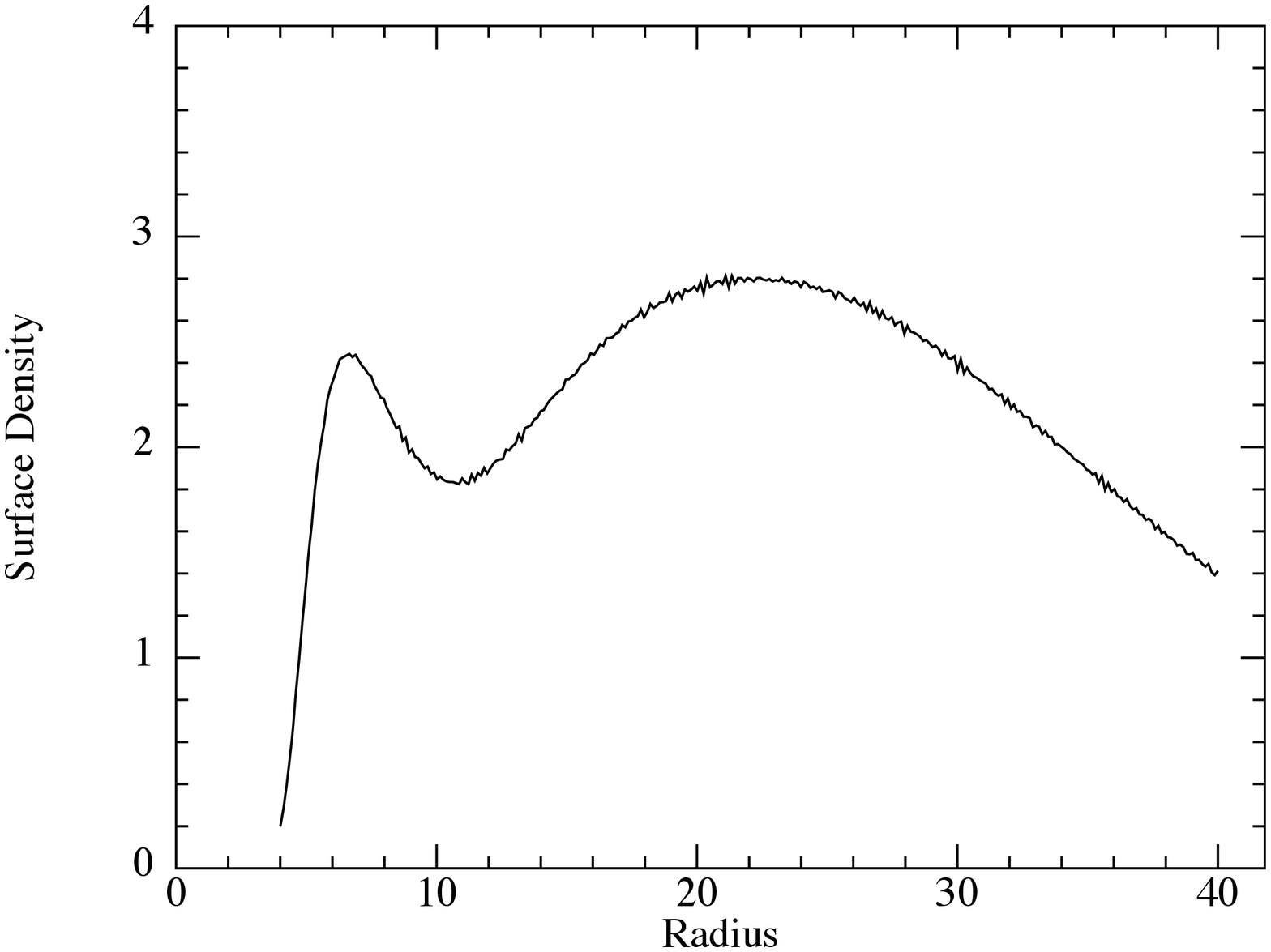} \hspace{1cm}
\includegraphics [width=0.9\columnwidth] {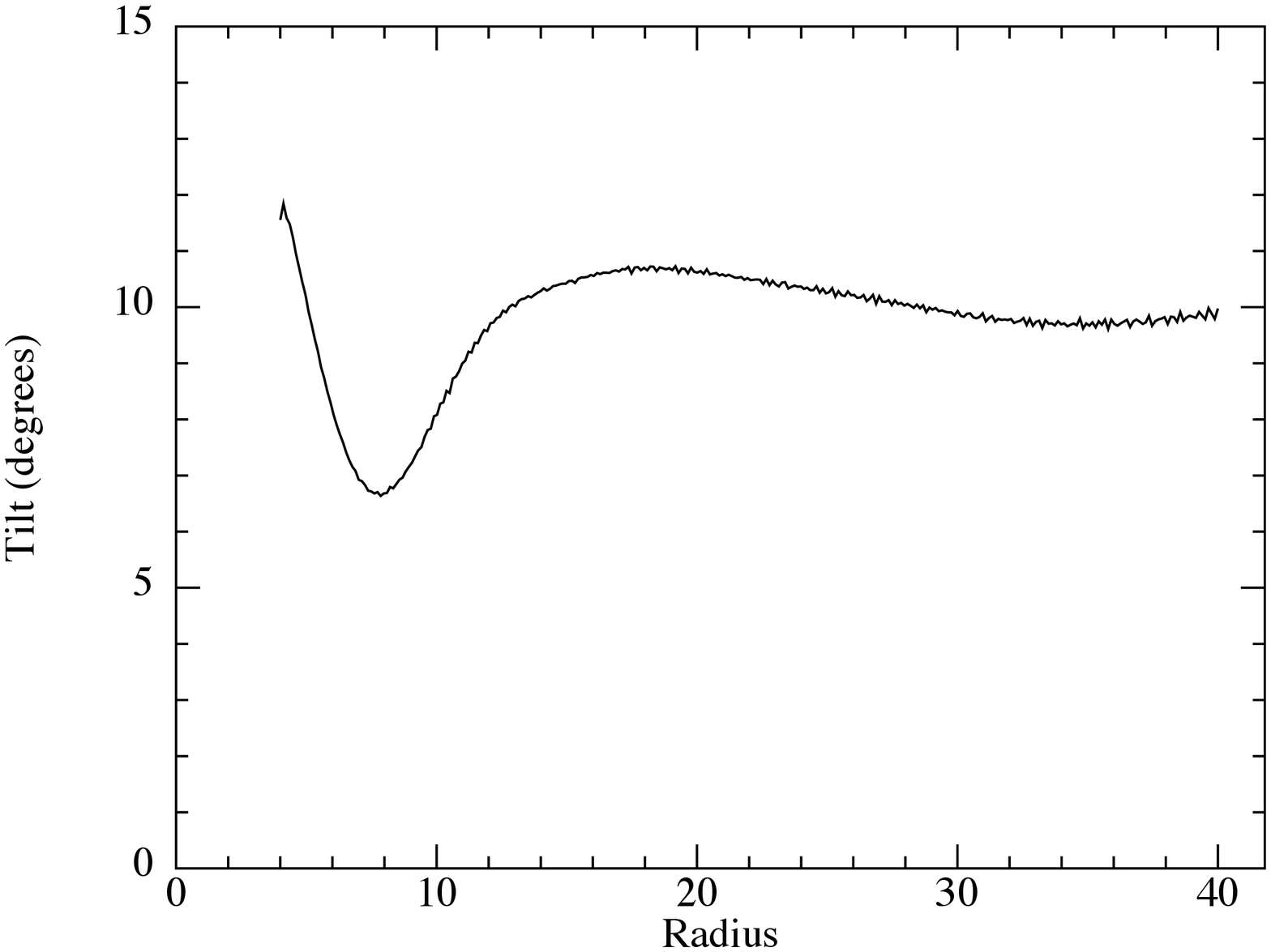}
\caption{Surface density (left) and tilt (right) after 25 orbits when apsidal precession is included.}
\label{fig:einstein_only}
\end{figure*}

\subsection{Apsidal precession neglected}
\label{subsection:neglected}
The right panel of Figure~\ref{fig:lt_only} shows the tilt as a function of radius for the simulation with a Keplerian potential, i.e. where apsidal precession is neglected. 
The start of the simulation is marked by a sharp disturbance in the tilt and surface density profiles as the black hole torque is applied, but this wave damps within $4.5$ orbits and the rest of the disc evolution is gradual (this initial disturbance occurs in all of the simulations reported). Our disc evolves in the same manner described by \citetalias{Krolik:2015lr}; the inner edge aligns with the spin of the black hole, and there is a smooth transition to the outer region which remains misaligned. The left of Figure~\ref{fig:lt_only} shows the  surface density profile of our disc. Similar to \citetalias{Krolik:2015lr}, there is little evolution of the surface density profile when apsidal precession is neglected. 
To allow direct comparison, we also show figures plotted in the same way as in \citetalias{Krolik:2015lr} such that Figures~\ref{fig:twist_lt}~and~\ref{fig:tilt_lt} are directly comparable to Figures 2 and 3 of their MHD simulation. The maximum values of the tilt and twist measured from our simulation are within $15\%$ of theirs and the evolution of our disc is in qualitative agreement. 
The comparison between our $\alpha$ -- disc simulations neglecting apsidal precession and the magnetohydrodynamical simulation of \citetalias{Krolik:2015lr} confirms that an $\alpha$ -- disc simulation gives similar results to an MHD one. That is, in at least this case MHD discs and pure $\alpha$ -- discs show equivalent results (\citealt{Sorathia:2013fk} found a similar result).

\begin{figure*}
\centering
\begin{minipage}{.47\textwidth}
  \centering
  \includegraphics[width=0.95\columnwidth]{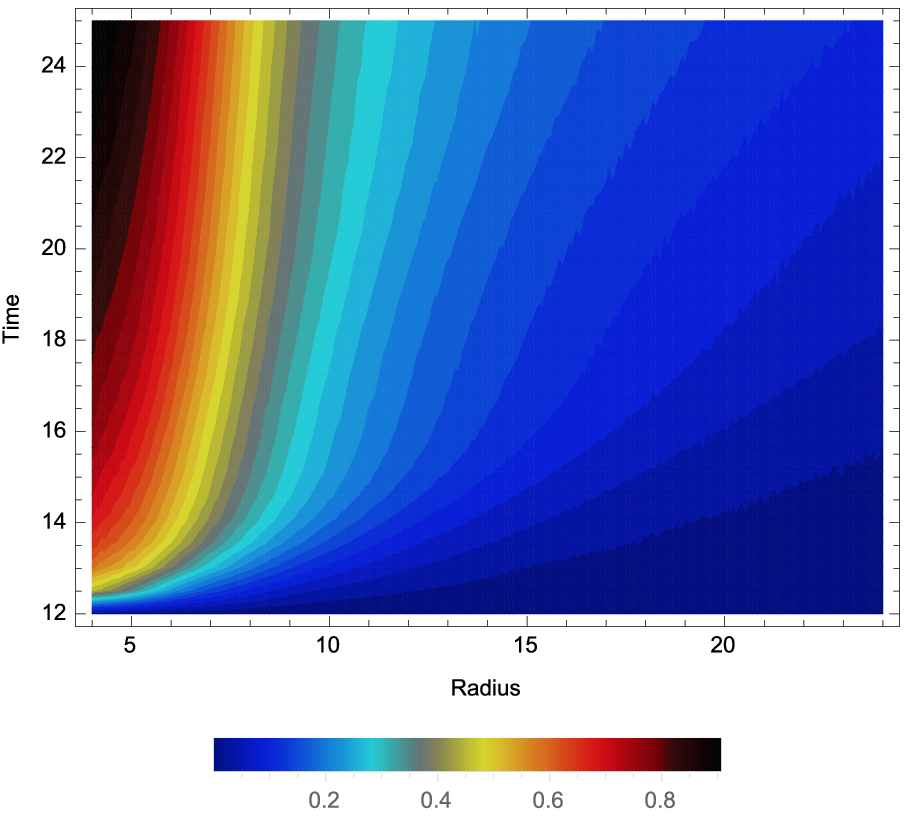}
  \captionof{figure}{Twist, measured in radians$/\pi$, as a function of time and radius in a disc including apsidal precession.}
  \label{fig:twist_einstein}
\end{minipage}\hfill
\begin{minipage}{.47\textwidth}
  \centering
  \includegraphics[width=0.95\columnwidth]{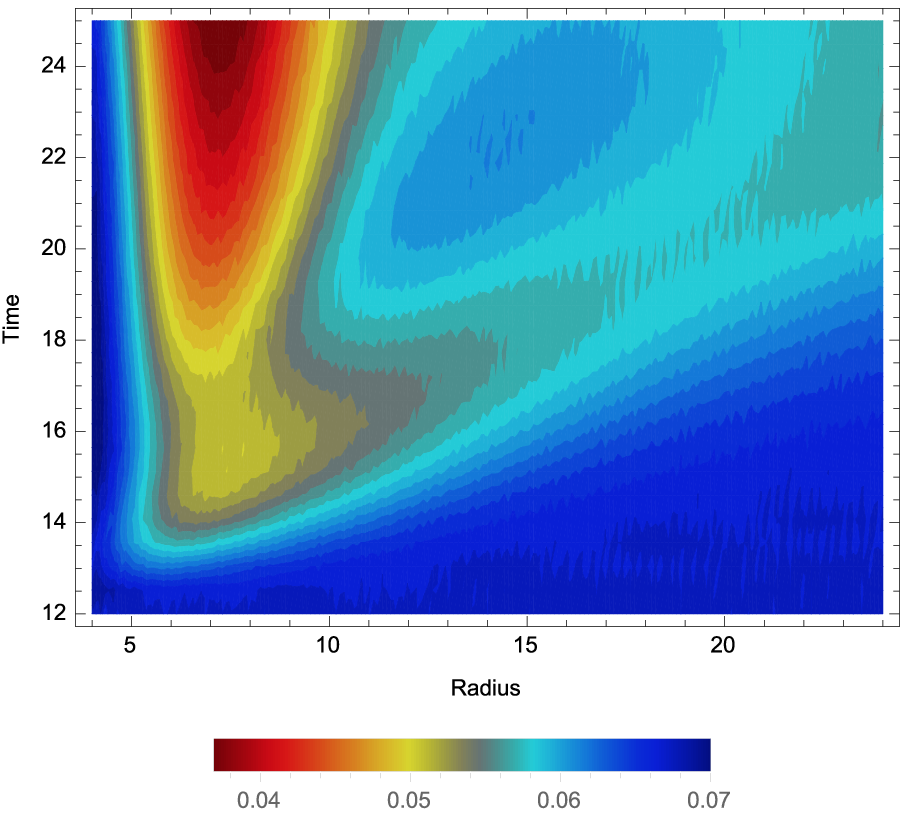}
  \captionof{figure}{Tilt, measured in radians$/\pi$, as a function of time and radius in a disc including apsidal precession.}
  \label{fig:tilt_einstein}
\end{minipage}
\end{figure*}

\subsection{Apsidal precession included}
Figures~\ref{fig:einstein_only},~\ref{fig:twist_einstein}~and~\ref{fig:tilt_einstein} show our results for discs with the same initial setup as Section~\ref{subsection:neglected}, but this time with the full effective GR potential, so that apsidal precession is explicitly present.
Here, in contrast to simulations which neglect it, we see that the inner edge of the disc remains misaligned. The tilt of the disc then decreases to a minimum around $R\sim9.5$ before increasing with radius.
%
%
The outer edge of this disc differs from that of the previous case (also seen in the tilt comparison), suggesting that a larger disc is required to prevent the outer boundary affecting the disc evolution.

The marked differences in the simulation results with and without apsidal precession show that including apsidal precession is crucial to give even qualitatively correct results.

%
%

\begin{figure*}
\includegraphics [width=0.98\textwidth] {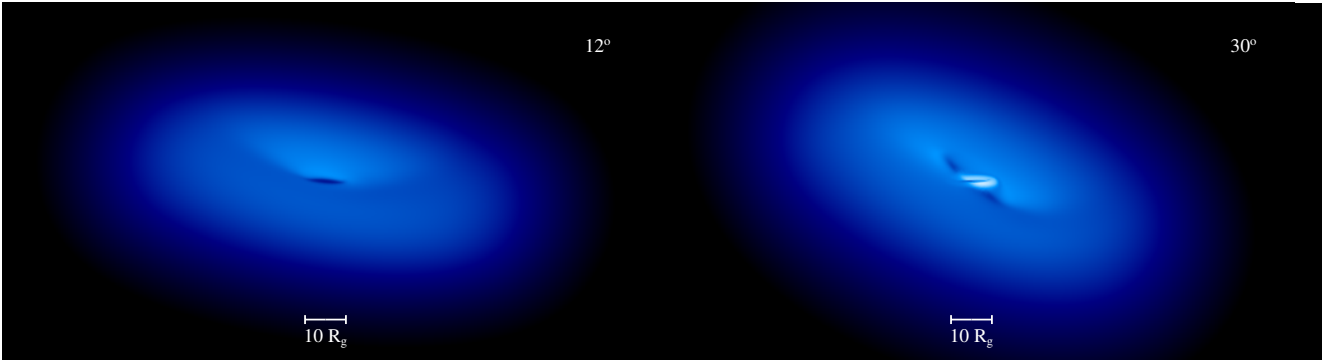}
\caption{Density rendering of warped accretion discs inclined at $12^{\circ}$ (left) and $30^{\circ}$ (right) to the black hole spin (spin axis is vertical in the plane of the page, i.e. along the $z$ axis). Evidence of disc breaking is seen only in the high inclination case.}
\label{fig:inclination_comparison}
\end{figure*}

\subsection{Breaking discs}

\cite{nixon_et_al_2012b} showed that one can estimate the radius at which a disc may break break by equating the Lense--Thirring precession torque with the internal viscous torque. This initial investigation neglected the additional viscous torque arising from a warp. Using this simplified criterion for $\beta=12^{\circ}$ would give $R_{break} \lesssim 20$, suggesting that the disc should have broken, contrary to the numerical results of both ourselves and \citetalias{Krolik:2015lr}. But this conclusion is wrong: neither of these simulations is in the diffusive regime where this criterion would apply. The relevant criterion for the wavelike regime involves a comparison of the sound crossing and precession timescales \citep{Nealon:2015fk} and shows that the disc should not break, as found in the simulations.\footnote{Even in the diffusive regime, \citet{Dogan:2015rt} showed that at small inclinations or when $\alpha$ is small, it is unreasonable to neglect vertical viscosity when estimating the internal torque in the disc. As these discs have low inclination and $\alpha\sim0.03$, it would be crucial to include the vertical torque in the breaking criterion if this simulation was diffusive \citep{Dogan:2015rt}.}

We address the question of whether the disc can break by simulating a disc with the same initial mass distribution, but this time with an inclination of $30^{\circ}$, again for $12.5$ orbits and using the effective GR potential. As the disc evolves the inner edge decreases from the initial $30^{\circ}$ but does not fully align with the black hole. The tilt then increases sharply around $R\sim10.5$, suggesting a break in the disc, before decreasing gradually at larger radii. Figure~\ref{fig:inclination_comparison} shows a density rendering of the $30^{\circ}$ disc at the end of the simulation, compared to the warped $12^{\circ}$ disc. For the high inclination disc, the separated innermost disc is only slightly misaligned, whilst the inner edge of the outer disc is inclined at $\sim30^{\circ}$.

\section{Discussion}
\label{section:discussion}
The striking differences that occur between simulations of warped wavelike discs that include or ignore the effects of apsidal precession confirm that it has a strong impact on the disc evolution. The inner edge of the disc is altered completely (from aligned to misaligned) and the oscillatory behaviour observed is consistent with previous analytical \citep{i_and_i_1997,lubow2002evolution} and numerical \citep{Nealon:2015fk} studies. We therefore conclude GR effects like this must be modelled accurately in simulations. This is achievable by either an appropriate post-Newtonian approximation, as used here \citep{Nealon:2015fk}, or a GR treatment \citep[e.g.][]{Fragile:2007uq,Morales-Teixeira:2014lr}.

The comparison between our hydrodynamic simulation and the MHD simulation of \citetalias{Krolik:2015lr} confirms that the disc dynamics can indeed be captured by hydrodynamics with an $\alpha$ viscosity. To date, \citet{Morales-Teixeira:2014lr} have completed the thinnest disc simulation with both GR and MHD taken into account, with an aspect ratio of $H/R\sim0.08$. However, this simulation was still in the bending wave regime as the viscosity parameter measured from the simulation was $\alpha\sim0.01$. In contrast, SPH simulations have already studied the diffusive regime, with $H/R\simeq0.01$ and $\alpha\simeq0.1$ \citep{nixon_et_al_2012}. GRMHD simulations to date have been unable to simulate discs which would have been expected to undergo disc breaking or tearing.
For the disc parameters used in the simulations presented in this paper, we show that disc breaking is possible when the disc is inclined at $\beta \gtrsim 30^{\circ}$. 

\section{Conclusion}
\label{section:conclusion}
We have conducted hydrodynamical simulations of accretion discs in the bending-wave regime with an  $\alpha$ viscosity. By comparing with the MHD simulation of \citetalias{Krolik:2015lr}, we confirmed that hydrodynamical simulations 
using an $\alpha$ viscosity capture the dominant evolution of warped accretion discs, showing results that are remarkably similar to the MHD simulations. 
%
%
We have shown that modelling the apsidal precession in the disc strongly affects its evolution. Simulations that do not take apsidal precession into account cannot give the correct disc evolution, which has nonzero disc tilt at the inner edge, and stable tilt oscillations with radius in the central disc regions. Finally, as expected, for the disc parameters chosen by \citetalias{Krolik:2015lr} we find no breaking, but demonstrate that a disc with the same parameters but a larger inclination does break.

\section*{Acknowledgments} 
RN is supported by an Australian Postgraduate Award. DJP is supported by an Australian Research Council Future Fellowship. CN thanks NASA for support via the Einstein Fellowship Programme, grant PF2-130098. CN was supported by the Science and Technology Facilities Council (grant number ST/M005917/1). Research in theoretical astrophysics at Leicester is supported by an STFC Consolidated Grant. We used the gSTAR national facility at Swinburne University, funded by Swinburne and the Australian Government's Education Investment Fund. We used \textsc{splash} \citep{Price:2007kx}.

\bibliographystyle{mnras} 
\bibliography{master}

\label{lastpage}
\end{document}